\documentclass[conference]{IEEEtran}
\usepackage{booktabs,multirow}
\usepackage{graphicx}
\usepackage{amsmath}
\usepackage{subeqnarray}
\usepackage{amsmath,amsthm,amsfonts,amssymb,bm}

\begin{document}
\bibliographystyle{IEEEtran}
\title{Characterizing Spatial Patterns of Base Stations in Cellular Networks}

\author{\IEEEauthorblockN{Qianlan Ying\IEEEauthorrefmark{1}\IEEEauthorrefmark{2},
Zhifeng Zhao\IEEEauthorrefmark{1}\IEEEauthorrefmark{2}, Yifan Zhou\IEEEauthorrefmark{1}\IEEEauthorrefmark{2}, 
Rongpeng Li\IEEEauthorrefmark{1}\IEEEauthorrefmark{2},
Xuan Zhou\IEEEauthorrefmark{1}\IEEEauthorrefmark{2}\IEEEauthorrefmark{3}, and
Honggang Zhang\IEEEauthorrefmark{1}\IEEEauthorrefmark{2}\IEEEauthorrefmark{3}}

\IEEEauthorblockA{\IEEEauthorrefmark{1}York-Zhejiang Lab for Cognitive Radio and Green Communications}

\IEEEauthorblockA{\IEEEauthorrefmark{2}Dept. of Information Science and Electronic Engineering\\Zhejiang University, Zheda Road 38, Hangzhou 310027, China\\Email: \{greenjelly, zhaozf, zhouyftt, lirongpeng, zhouxuan, honggangzhang\}@zju.edu.cn}

\IEEEauthorblockA{\IEEEauthorrefmark{3}Universit\'{e} Europ\'{e}enne de Bretagne \& Sup\'{e}lec, Avenue de la Boulaie, CS 47601, 35576 Cesson-S\'{e}vign\'{e}  Cedex, France\\Email: Honggang.Zhang@supelec.fr}
}
 
\maketitle
\begin{abstract}
\boldmath 
The topology of base stations (BSs) in cellular networks, serving as a basis of networking performance analysis, is considered to be obviously distinctive with the traditional hexagonal grid or square lattice model, thus stimulating a fundamental rethinking. Recently, stochastic geometry based models, especially the \textit{Poisson point process} (PPP), have been attracting an ever-increasing popularity in modeling BS deployment of cellular networks due to its merits of tractability and capability for capturing non-uniformity. In this study, a detailed comparison between common stochastic models and real BS locations is performed. Results indicate that the PPP fails to precisely characterize either urban or rural BS deployment. Furthermore, the topology of real data in both regions are examined and distinguished by statistical methods according to the point interaction trends they exhibit. By comparing the corresponding real data with aggregative point process models as well as repulsive point process models, we verify that the capacity-centric BS deployment in urban areas can be modeled by typical aggregative processes such as the \textit{Matern cluster process}, while the coverage-centric BS deployment in rural areas can be modeled by representative repulsive processes such as the \textit{Strauss hard-core process}.
\end{abstract}
\section{Introduction}
\IEEEPARstart{T}{he} topological structure of cellular networks has been gaining tremendous complexity to fulfill exponentially increasing demand for mobile data. Nowadays practical deployment of cellular networks, which are organically deployed to provide high capacity, is considered to be highly heterogeneous and irregular \cite{andrews2010primer}. 

Since the architecture of heterogeneous networks plays a key role in evaluating system performance in 4G and future cellular networks, analysis of the networking topology is emerging as a primary task for subsequent accurate performance characterization. By far, the most common assumption widely used in analytical calculations for cellular networks is that base stations (BSs) are uniformly distributed in the covered areas. Accordingly, hexagonal grids or square lattices are pervasively utilized to model the locations of BSs. Both models, however, are generally intractable and structurally different from the real BS deployment. 

Therefore, much of the researchers' focus has been shifting to more accurate BSs' spatial characterization, so as to capture the non-uniformity of practical deployment. In that regard, stochastic geometry has proven to be an effective means to model BS placement \cite{haenggi2009stochastic}. It is demonstrated in \cite{andrews2011tractable} that the \textit{Poisson point process} (PPP), of which the points are independently and uniformly distributed in certain area, tracks the real configurations as accurately as the conventional grid model. Particularly, it is applicable in deriving tractable theoretical results for downlink performance evaluation under some assumptions. However, as population usually distributes unevenly and practical BSs tend to be deployed with a target (e.g., coverage-centric or capacity-centric) being strongly associated with human activities, the PPP exhibits some unrealistic characteristics. Conversely, alternative spatial patterns might yield superior modeling precision. For example, researchers have showed the fitness of the \textit{Geyer saturation process} in modeling the spatial patterns of WiFi spots \cite{riihijarvi2010modeling}. For cellular networks, Taylor et al. proposed to use the \textit{Strauss process} and the \textit{Geyer saturation process} to model macro-cellular BS deployment \cite{taylor2012pairwise}. The study in \cite{lee2013stochastic} indicated the accuracy of Poisson cluster models in characterizing BS spatial distributions in large cities. Guo and Haenggi analyzed the fitness of the Strauss process using the urban and rural data collected from a public dataset in UK \cite{guo2013spatial}. But their study was lacking of in-depth consideration on the difference between point patterns in various kinds of regions limited by the relatively small BS dataset. Thus, although there exist several works towards spatially modeling the BS locations by stochastic geometry tools in different scenarios, few of them has ever shed light on the essential differences between the two typical geographical categories, i.e., urban and rural areas.

Motivated by the observations above, we try to find the most precise models for BSs in different types of geographical areas accordingly, based on a large BS dataset collected from a cellular network operator in China. Firstly, we select data subsets from the urban area and the rural area separately and demonstrate that the PPP model is pessimistic in modeling either point pattern. Specifically, by means of summary statistics including the $G$-function, $K$-function and $L$-function \cite{ripley1977modelling} \cite{baddeley2008analysing}, we measure the spatial dependence of both kinds of regions with a large number of the measured data. Furthermore, we test the hypotheses of different spatial models by the $L$-function and the coverage probability, and verify the accuracy of several stochastic models matching urban and rural BS deployment, respectively.
    
The rest of the paper is organized as follows. In Section \uppercase\expandafter{\romannumeral 2}, we begin with a brief overview of representative stochastic spatial models and present a description of the BS dataset. Then, the methodology of fitting point processes to the collected practical data is introduced in Section \uppercase\expandafter{\romannumeral 3}. The experimental results are shown in Section \uppercase\expandafter{\romannumeral 4} before concluding this study in Section \uppercase\expandafter{\romannumeral 5}.

\section{Preliminaries}
\subsection{Background on Point Processes}
Generally, a point process $\mathbf{x}$ is a finite collection of randomly distributed locations contained in a given bounded region $S$. A realization of such a point process ${\bm{\{}}{x_1},{x_2},...,{x_{N(\mathbf{x})}}\}$ can be specified by the number of points $N(\mathbf{x})$ and the joint distribution of the points in $\mathbf{x}$. Point processes (e.g., pairwise interaction processes, hard-core processes and clustered processes) can be grouped into three categories, the PPP, repulsive processes and aggregative processes. A PPP $\mathbf{x}$ can be defined on $S$ with its intensity measure $\lambda$ satisfying $\lambda(B)>0$ for any bounded region $B$ belonging to $S$, while $N(B)$ is Poisson distributed with mean $\lambda(B)$. They possess the property of ``no interaction'' between nodes \cite{moller2007modern}. 

By contrast, a pairwise interaction process $\mathbf{x}$ takes inter-point interactions into consideration. Its probability density function (PDF) on a compact region satisfies
\begin{equation}
f(\mathbf{x}) = a \cdot \prod\limits_{i = 1}^{N(\mathbf{x})} {b({x_i}) \cdot \prod\limits_{i < j} {c(} {x_i},{x_j})} ,
\end{equation}
where $a$ denotes a normalizing constant, while $b( \cdot )$ and $c( \cdot )$ are nonnegative functions, indicating first-order trends and pairwise interactions, respectively. Usually, when the functions $b( \cdot )$ and $c( \cdot )$ take different forms, the pairwise interaction process could be simplified to different processes:

\begin{itemize}

\item{\bf\textit{Strauss process:}} $b( \cdot )=\beta$ and $c(u,v)=\gamma $ conditional on $0 < \left\| {u - v} \right\| \le r$, $c(u,v) = 1$ otherwise. Thus, we have the density function as the form 
\begin{equation}
f(\mathbf{x}) = a{\beta ^{N(\mathbf{x})}}{\gamma ^{s(\mathbf{x})}} , 
\end{equation}
in which $s(\mathbf{x})$ denotes the number of point pairs of $\mathbf{x}$ lie within a distance $r$. The \textit{Strauss process} is effective in modeling repulsion effect between points, yet it proves to be non-integrable for $\gamma  > 1$ corresponding to the desired clustering.

\item{\bf\textit{Geyer saturation process:}} As a generalized version of \textit{Strauss process}, a saturation limit $sat$ is added in the exponent of $\gamma$, thereby trimming the total contribution from each point's pairwise interaction to a maximum $sat$. The process can describe both repulsive and aggregative patterns.

\end{itemize}
 
Hard-core processes (e.g., the Poisson hard-core process and the Strauss hard-core process), implied by the name, introduce a hard-core distance $h_c>0$ into its PDF \cite{diggle1994parameter}. Therefore, the inter-point distances between distinct points are always greater than ${h_c}$.

On the other hand, cluster processes can precisely fit point patterns with the aggregation behavior. In this study, we primarily focus on the \textit{Matern cluster process} (MCP), which is a special case of the Poisson cluster process. Usually, The Poisson cluster process is formed by taking a Poisson process as parent points with daughter points scattering around. In particular, we call it the MCP if the daughter points are uniformly distributed within the ball of radius $r$ around their parent points. An MCP can be further specified by other parameters, including the intensity of the parent points $\kappa$ and the mean number of points in each cluster $\mu$.
\subsection{Dataset Description}
Our dataset is based on the real measurement of commercial base stations in an eastern province in China, which includes 47663 BS location records with more than 40 million subscribers involved. The whole research area covers 101,800 ${\rm{k}}{{\rm{m}}^2}$ and is generally divided into two typical regions, namely the urban area and the rural area by matching the longitude and latitude information of each BS to that of the target area on a Google Map. 

\begin{figure}[htbp]
\setlength{\abovecaptionskip}{0pt}
\setlength{\belowcaptionskip}{0pt}
\centering
\includegraphics[clip, trim=10mm 0mm 3mm 0mm, height=3.7in, width=3.5in]{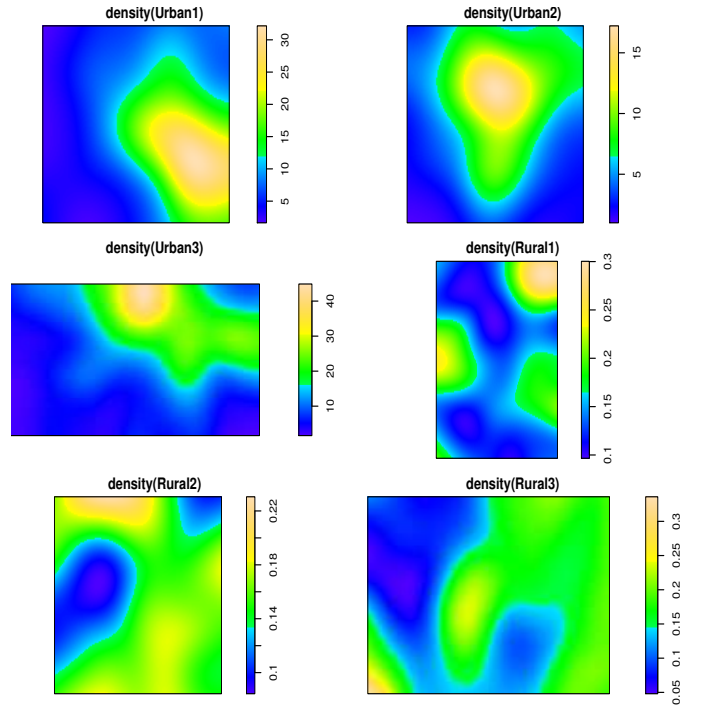}\\ 
\caption{BS kernel density estimates of representative urban and rural areas.}\label{fig0}
\end{figure}

On\textit{} the macroscopic level, the BS density varies heavily across densely populated areas and less populated areas, whose difference can be easily explained by the respective capacity requirements. This phenomenon can be better illustrated in Fig. 1 where we randomly pick three urban areas and three rural areas from the collected data and draw the kernel density estimate for each of them. 

To ensure the data richness, the selected areas $Urban 1$ and $Urban 2$ are both selected from inland cities while $Urban 3$ is from a coastal city. On the other hand, the three rural areas $Rural 1$, $Rural 2$ and $Rural 3$ are randomly selected from three different less populated areas without overlapping.  As seen in Fig. 1, the shape of the color temperature maps is varied with BS density. The different variation ranges between subfigures show that the average number of BSs per square kilometer in the urban areas is 10 to 100 times larger than that in the rural areas, which also confirms the region imbalance on BS spatial distribution. We will conduct more detailed statistical tests using these regions in Section \uppercase\expandafter{\romannumeral 4}.


In order to identify pairwise interaction trends between points by the corresponding metrics, we mainly focus on the representative areas of both kinds in relatively compact datasets and ignore the density difference by mapping the selected point patterns onto the same scale in the following study. Thus, we pick a square region $U$ with a total space of $7.73 k{m^2}$ from a dense urban region in the provincial capital city and a square region $R$ covering a total area of $1069.89 k{m^2}$ in typical part of the rural area. Both regions are mapped onto unit squares as shown in Fig. 2.

\begin{figure}[htbp]
\centering
\includegraphics[clip,width=3.2in]{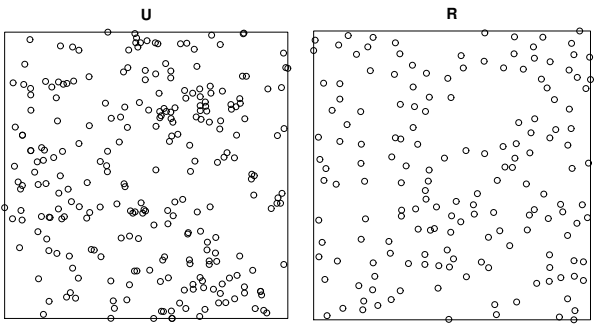}\\
\caption{\textit{(left)}: Distribution of the urban point pattern $U$ with 259 BSs. \textit{(right)}: Distribution of the rural point pattern $R$ with 167 BSs.}\label{fig1}
\end{figure}

\section{Methodology}
This paper aims at obtaining an accurate point process to model the real BS deployment in target areas. In order to reach such a goal, we utilizes the following fitting and analysis methods.

\subsection{Fitting Method}
It is common to utilize the maximum pseudolikelihood estimates as the fitting method in stochastic geometry. Whereas for models with irregular parameters such as $sat$ in the \textit{Geyer saturation process}, we can obtain the relevant parameters by analogy with the profile maximum pseudolikelihood estimator \cite{baddeley2000practical}.


\subsection{Analysis Method}
The analysis of point patterns largely depends on summary statistics (e.g., the standard deviation). For testing our assumption of real data, we employ the nearest neighbor distance function (the $G$-function), the $L$-function and the coverage probability as the metrics:
\subsubsection{$G$-function}
The $G$-function is the cumulative frequency distribution of the nearest neighbor distances, which illustrates the ways the points are spaced. It indicates clustering when the $G$-function increases rapidly at a short distance, while it implies dispersion otherwise.

\subsubsection{$L$-function}
The $L$-function relies on the distance between the points and provides an estimate of spatial dependence over a wide range of scales. Therefore, the $L$-function could judge whether a pattern exhibits clustering or dispersion. Formally, the $L$-function is defined as $L(r)=\sqrt {{{K(r)} \mathord{\left/ {\vphantom {{K(r)} \pi }} \right. \kern-\nulldelimiterspace} \pi }} $. Here, $K(r)$ is the Ripley's $K$-function and can be calculated as 
\begin{equation}
K(r) = \frac{1}{\lambda }\mathbb{E}[N(\mathbf{x} \cap b(x,r)\backslash \{ x\} )| x \in \mathbf{x}],
\end{equation}
where $\lambda$ denotes the intensity of the points \cite{ripley1977modelling}. If a point pattern is the PPP, $L(r)$ would equal to $r$. Moreover, $L(r)>r$ indicates clustering while $L(r)<r$ represents dispersion.

\subsubsection{Coverage Probability}
The coverage probability is another important metric related to network performance. It indicates the probability that a randomly chosen mobile user connected to the nearest BS achieves signal-to-interference-plus-noise-ratio (SINR) larger than a threshold $T$ \cite{andrews2011tractable}. Specifically, we assume all the BSs with one antenna transmits with the equal power $P$, and a typical user receives the signal power from the nearest BS with the location ${x_k} \in \mathbf{X}$ and the interference power as the total received power from all the other BSs. Then the SINR of the typical user located at the origin should be written as the form:
\begin{equation}
SINR = \frac{{P{h_k}{{\left\| {{x_k}} \right\|}^{ - \alpha }}}}{{W + {\sum _{i:{x_i} \in {\bf{x}}\backslash {x_k}}}P{h_i}{{\left\| {{x_i}} \right\|}^{ - \alpha }}}},
\end{equation}
where $W$ denotes the noise power, $h_i \in \mathbf{H}$ is the fading and shadowing coefficient, and ${{{\left\| {{x}} \right\|}^{ - \alpha }}}$ is the standard path loss between the BS located at ${x}$ and the user. Here Rayleigh fading is exponentially distributed with mean 1 and lognormal shadowing is of the value ${10^{X/10}}$ $(X$$\sim$$N(0,\sigma _{LN}^2))$. The path loss coefficient satisfies  ${\alpha} > 2$ for testing as the outdoor scenario without losing generality.

\section{Experimental Results and Analyses}
\subsection{Pre-judgement}
First we examine the $G$-function and the $K$-function of $U$ and $R$ to determine whether they are clustered or repulsive. The pre-judgement results will help to understand both point patterns and find appropriate spatial models.

\begin{figure}[htbp]
\setlength{\abovecaptionskip}{0pt}
\setlength{\belowcaptionskip}{0pt}
\centering
\includegraphics[clip,width=3.5in]{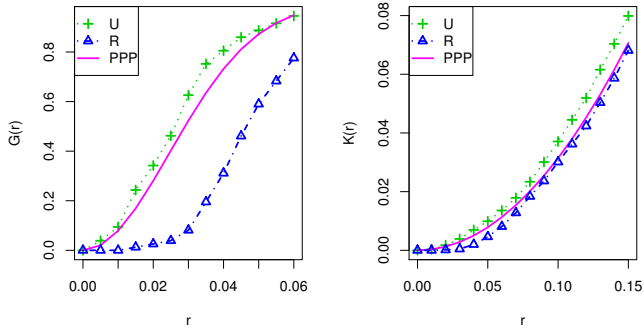}\\
\caption{\textit{(left)}: The $G$-function of $U$, $R$ and the theoretical $G$-function of the PPP (the solid line).  \textit{(right)}: The Ripley's $K$-function of $U$, $R$ and the theoretical $K$-function of the PPP.}\label{fig2}
\end{figure}

As seen in Fig. 3, the point pattern $U$ in dense urban area has a strong tendency of clustering between BSs with both calculation results larger than the theoretical ones (i.e. the PPP). On the contrary, the point pattern $R$ reflects diffusive regularity as the corresponding curve in $G$-function increases much lower at a short distance. 

We next utilize the similar metric for the datasets in Fig. 1 to show the result above is more universal. For each of the six regions, we firstly generate a total number of 10000 square subregions at random locations and ensure the BS number inside is suitable for the analysis in R software. We set the range of BS numbers as 60 to 220. Based on the $L$-function, a point pattern in the subregion is characterized as clustering only if $L(r)\ge r$ can be satisfied and repulsive only if $L(r)\le  r$ is satisfied for every $r$ belongs to a normalized interval in our validation. We set the interval as $(0,0.15]$ here.

\begin{table}[htbp]
  \centering
  \caption{The probability of clustering and repulsion for the urban and rural areas}
    \begin{tabular}{ccccc}
    \toprule
          & Point Number & Area($k{m^2}$)  & Clustering & Repulsive \\
    \midrule
    Urban1 & 2349  & 210.86 & 66.91\% & 0.03\% \\
    Urban2 & 1468  & 225.07 & 87.74\% & 0.01\% \\
    Urban3 & 1001  & 83.54 & 79.71\% & 0.04\% \\
    Rural1 & 318   & 2035.9 & 1.01\% & 54.53\% \\
    Rural2 & 623   & 4141.3 & 0.31\% & 58.41\% \\
    Rural3 & 288   & 1962.9 & 0.19\% & 16.16\% \\
    \bottomrule
    \end{tabular}%
  \label{tab:addlabel}%
\end{table}%

We show the clustering and repulsive probability of the six regions in the columns of Table I. Compared with the repulsive probability, the clustering probability reaches significant levels for all the three urban regions. For the rural regions, the results are somewhat the reverse of those in the urban regions, as the tendency of repulsion is obvious except in the point pattern of $Rural 3$ which however, also indicates low probability of clustering. 

These results above can be readily acknowledged, as network coverage optimization and interference minimization contribute to spatial regularity of BS deployment in rural regions. Whereas in densely populated regions, the comparatively higher amount of network traffic in urban areas contribute to intensive and clustering BS deployment.

\begin{table}[htbp]
  \centering
  \caption{Parameters for best model fitting for $U$ and $R$}
    \begin{tabular}{lll}
    \toprule
    Data type & Processes & Parameters \\
    \midrule
    \multirow{3}[0]{*}{Urban} & PPP & $\lambda=47.50$ \\
          & Geyer & $r=0.03$, $sat=4$, $\beta=182.93$, $\gamma=1.25$ \\
          & MCP & $\kappa=162.48$, $r=0.067$, $\mu=1.61$ \\
    \midrule
    \multirow{4}[0]{*}{Rural} & PPP & $\lambda=35.75$ \\
          & PHCP & ${h_c}=0.015$, $\beta=173.34$ \\
          & SH & ${h_c}=0.015$, $\beta=237.24$, $r=0.03$ \\
          & Geyer & $r=0.073$, $sat=1$, $\beta=26.08$, $\gamma=6.01$ \\
    \bottomrule
    \end{tabular}%
  \label{tab:addlabel}%
\end{table}%

\subsection{Fitted Models}
Inspired by the first-stage results above, we further examine the fitness of aggregative processes to the clustered point pattern $U$. The MCP and the \textit{Geyer saturation process} are included in the fitting process. On the other hand, repulsive processes including the \textit{Strauss hard-core process} (SH), the \textit{Geyer saturation process} and the \textit{Poisson hard-core process} (PHCP) are utilized to fit the more regular point pattern $R$. As a benchmark, the PPP is employed in both validations, which could further verify our hypotheses. Table II lists the corresponding fitting results, in terms of the maximum pseudolikelihood estimates and the profile maximum pseudolikelihood estimates. 

We use the R package \textit{spatstat} \cite{baddeley2008analysing} to fit parametric models to the real data, and then generate simulated realizations using Markov Chain Monte Carlo tests. 

\subsection{Model Validation}
After the parameter estimation, the fitting accuracy of these models above are assessed based on the $L$-function and the coverage probability separately. For each of the fitted models, we generate 599\footnote{The significance level $\alpha$ can be calculated as $\alpha = 2 * nrank/(1 + nsim)$, by the number of simulations $nsim$ and the rank of the envelope value $nrank$. Therefore, if we take $nsim=599, nrank=30$, $\alpha$ would equal $0.1$.} simulated curves and throw out 30 highest and 30 lowest values to form the pointwise envelope, which leads to a 90\% confidence level. Then we judge whether the $L$-function and the coverage probability of the realistic point patterns lie within these confidence intervals.

The coverage probability curve is drawn under a wide range of SINR thresholds. Notably, in order to eliminate the edge effect induced by user locations, 1000 randomly selected mobile users are assumed to be located in the central part of the unit window covering  $\frac{2}{3}$width$\times\frac{2}{3}$height of the total area. The coverage probability is then computed by comparing the corresponding average SINR values to the selected threshold.

\begin{figure}[htbp]
\setlength{\abovecaptionskip}{0pt}
\setlength{\belowcaptionskip}{0pt}
\centering
\includegraphics[clip,trim=5.5mm 0 12.2mm 0, width=3.2in,height=4.8in]{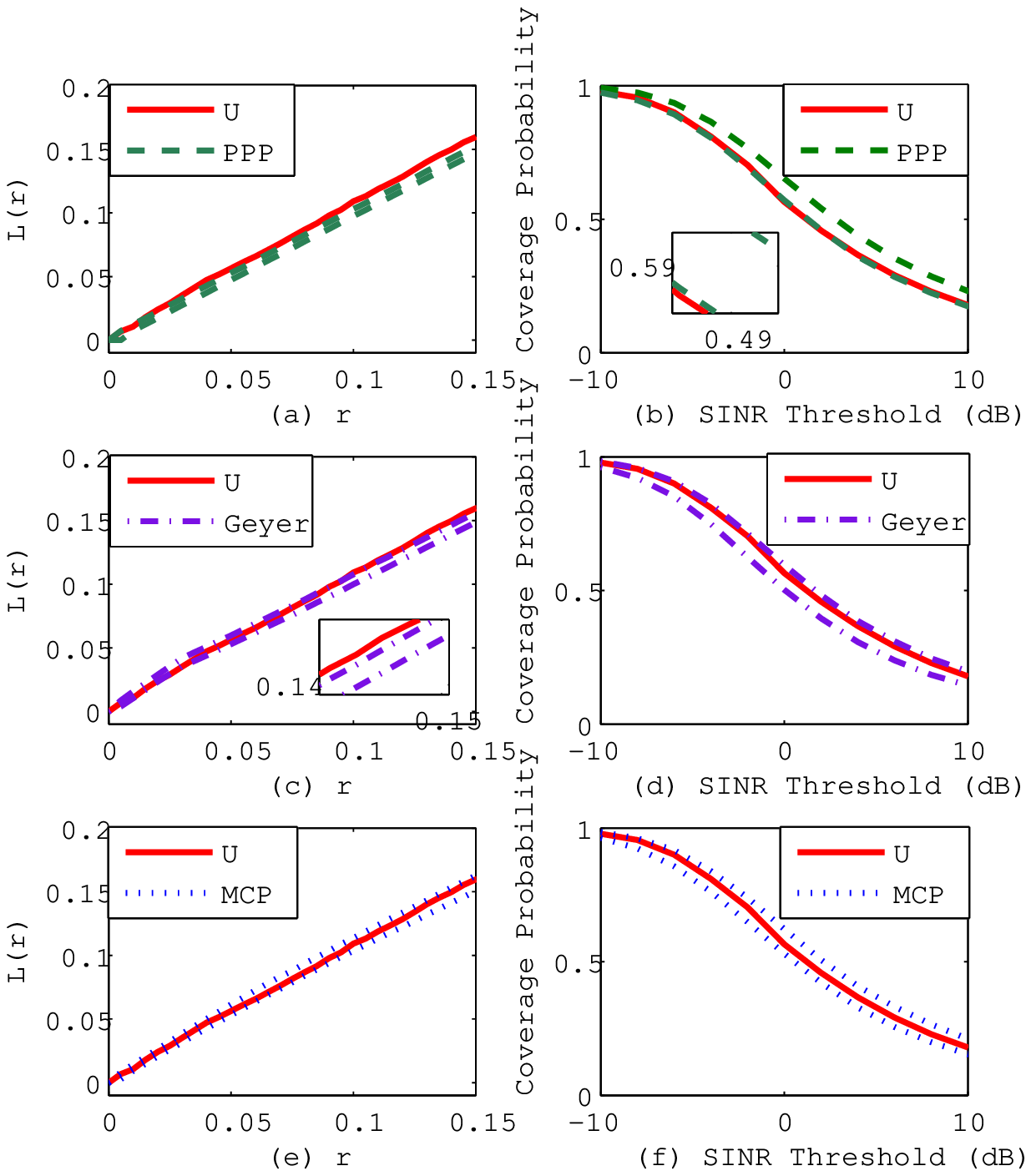}\\
\caption{\textit{(a), (c) and (e)}: the $L$-function of the point pattern $U$ and the corresponding envelopes of fitted models; \textit{(b), (d) and (f)}: the coverage probability function of the point pattern $U$ and the corresponding envelopes of fitted models.}\label{fig3}
\end{figure}

\begin{figure}[htbp]
\setlength{\abovecaptionskip}{-15pt}
\setlength{\belowcaptionskip}{0pt}
\centering
\includegraphics[clip, trim=5.5mm 0 12.1mm 5mm, width=3.2in, height=6.8in]{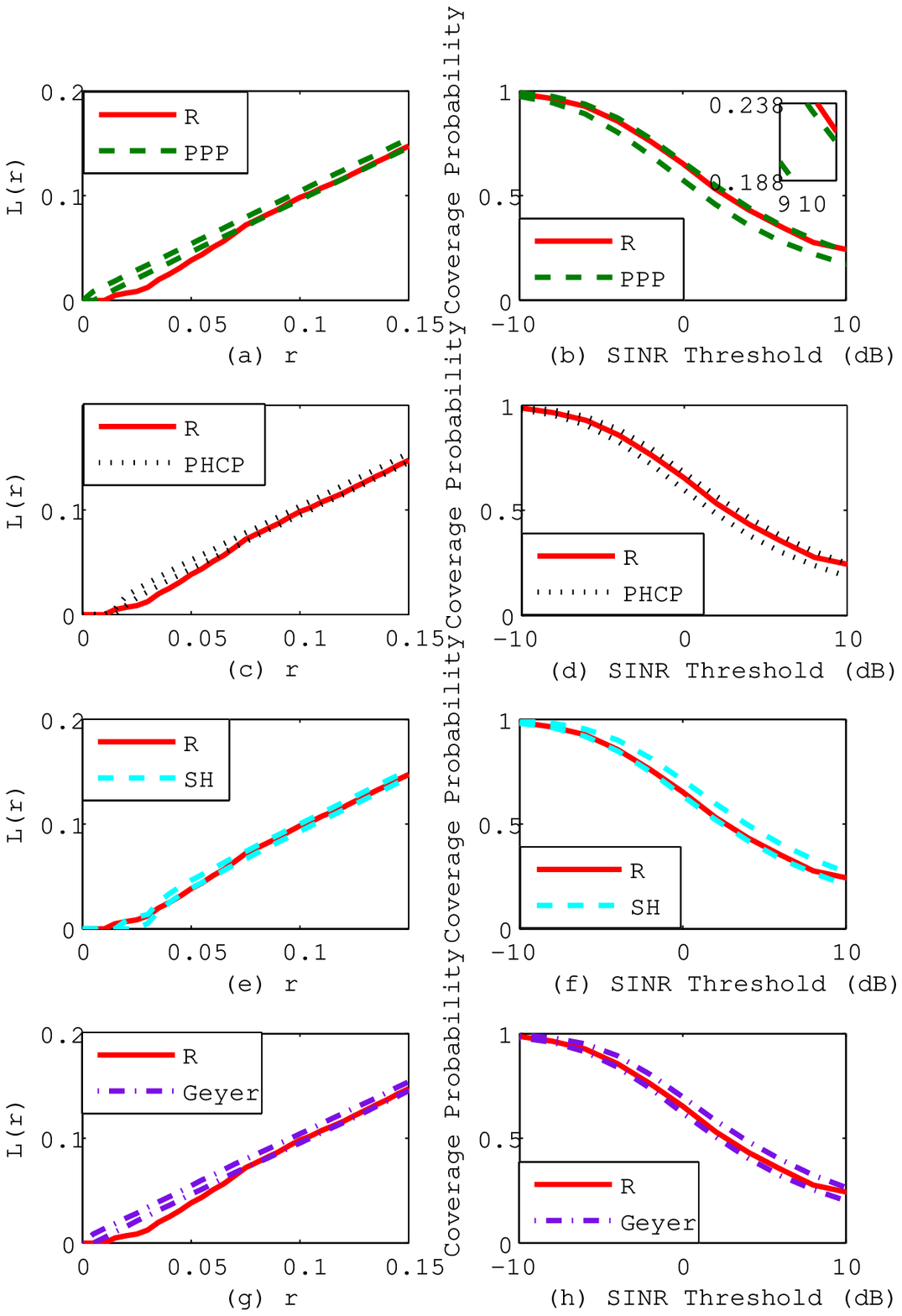}\\
\caption{\textit{(a), (c), (e) and (g)}: the $L$-function of the point pattern $R$ and the corresponding envelopes of fitted models; \textit{(b), (d) (f) and (h)}: the coverage probability function of the point pattern $R$ and the corresponding envelopes of fitted models.}\label{fig4}
\end{figure}

\subsubsection{Urban Point Pattern}
Fig. 4 presents the fitting envelopes with the 90\% confidence level of the PPP, the \textit{Geyer saturation process} and the MCP with respect to the urban point pattern $U$. The $L$-function curves are shown in Fig. 4 (a), (c) and (e) and the coverage probability curves are plotted in the other subfigures, respectively. Obviously, we can reject the hypothesis that $U$ is a PPP, as the observed function values on both metrics fall outside the envelope from Fig. 4 (a) and (b). According to Fig. 4 (d), we can not reject the hypothesis that $U$ is a \textit{Geyer Saturation process} by the coverage probability. However, in Fig. 4 (c), the curve of $U$ lies outside the envelope when $r > 0.1$. Therefore, we can also reject the null hypothesis of the \textit{Geyer} model. As seen in Fig. 4 (e) and (f), the MCP fits precisely to the point pattern $U$ based on both statistics. 

\textit{Remark}: The MCP is the model that captures the properties of the urban point pattern best.

\subsubsection{Rural Point Pattern}
Fig. 5 shows the realistic $L$-function and the coverage probability of the rural point pattern $R$, as well as the corresponding simulated envelopes, including the PPP, the PHCP, the SH and the \textit{Geyer saturation process}. We can reject the null hypothesis of the PPP using either the $L$-function or the coverage probability, as seen in Fig. 5 (a) and (b), respectively. Though the PHCP and the \textit{Geyer saturation process} satisfy the verifications based on the coverage probability, as depicted in Fig. 5 (d) and (h), we can reject these two hypotheses by Fig. 5 (c) and (g). However, we can not reject the hypothesis that $R$ is the SH on both metrics according to Fig. 5 (e) and (f).

\textit{Remark}: The SH fits best to the original rural pattern. 

\section{Conclusion}
In this paper, we investigated the problem of spatially modeling BS deployments based on the real data from practical deployment, and rejected the hypothesis that BS in either urban or rural area is PPP distributed. Afterwards, the Ripley's $K$-function and the $G$-function were utilized to reveal the discrepancy between the urban and the rural BS deployment. Furthermore, we found the urban BS deployment expresses clustering characteristics and could be well fitted by aggregative processes such as the \textit{Matern cluster process}, while the rural BS deployment implies dispersion phenomenon and follows a \textit{Strauss hard-core process}. These results indicate the diversity of BS deployment, and thus provide effective performance analysis approach, as well as quantized reference for cellular network planning.

\section*{Acknowledgement}
This study is supported by the National Basic Research Program of China (973Green, No. 2012CB316000) and the grant of ``Investing for the Future" program of France ANR to the CominLabs excellence laboratory (ANR-10-LABX-07-01)."
\bibliographystyle{IEEEtran}
\bibliography{main}

\end{document}